\documentclass[12pt]{iopart}
\usepackage{iopams}
\usepackage{amsmath2}
\usepackage{graphicx}
\usepackage{color}
\usepackage{amssymb}

\newcommand{\up}{\uparrow}
\newcommand{\dw}{\downarrow}

\newcommand{\anticomm}[2]{\left\{#1,#2\right\}}

\newcommand{\ii}{\text{i}}

\begin{document}

\title[Local spectral properties of Luttinger liquids]
{Local spectral properties of Luttinger liquids: scaling versus 
nonuniversal energy scales}

\author{D Schuricht$^1$, S Andergassen$^{1,2}$ and V Meden$^1$}

\address{$^1$Institut f\"ur Theorie der Statistischen Physik, RWTH Aachen University 
and JARA---Fundamentals of Future Information Technology, 52056 Aachen, Germany}

\address{$^2$Faculty of Physics, University of Vienna, Boltzmanngasse 5, 1090 Wien, Austria}

\ead{meden@physik.rwth-aachen.de}

\begin{abstract}
Motivated by recent scanning tunneling and photoemission 
spectroscopy measurements on self-organized gold chains on a germanium 
surface we reinvestigate the local single-particle spectral properties 
of Luttinger liquids. In the first part we use the bosonization approach 
to exactly compute the local spectral function of a simplified field theoretical 
low-energy model and take a closer look at scaling properties as a function 
of the ratio of energy and temperature. Translational invariant Luttinger 
liquids as well as those with an open boundary (cut chain geometry) are 
considered. We explicitly show that the scaling functions of both 
setups have the same analytic form. The scaling behavior suggests a variety 
of consistency checks which can be performed on measured data to 
experimentally verify Luttinger liquid behavior. 
In a second part we approximately compute the local 
spectral function of a microscopic lattice model---the extended 
Hubbard model---close to an open boundary using the functional 
renormalization group. We show that as a function of energy and temperature  
it follows the field theoretical prediction in the low-energy regime
and point out the importance of nonuniversal 
energy scales inherent to any microscopic model. 
The spatial dependence of this spectral function is characterized 
by oscillatory behavior and an envelope function which follows 
a power law both in accordance with the field theoretical continuum 
model. Interestingly, for the lattice model we find a phase shift
which is proportional to the two-particle interaction and not 
accounted for in the standard bosonization approach to Luttinger liquids 
with an open boundary. We briefly comment on the 
effects of several one-dimensional branches cutting the Fermi energy and 
Rashba spin-orbit interaction.  
\end{abstract}


\maketitle

\section{Introduction}
\label{sec:introduction}

Over decades theoretical studies of the single-particle spectral properties
of metallic one-dimensional (1d) correlated electron systems---so-called Luttinger 
liquids (LLs)---were ahead of the experimental attempts to find or synthesize 
appropriate quasi 1d materials and perform spectroscopy on them. In fact, while 
at the begining of the 1990's a clear picture of the basic spectroscopic 
properties of translational invariant LLs was established (for reviews 
see e.g.\ Refs.~\cite{Voit95}, \cite{Giamarchi03} and \cite{Schoenhammer05}) 
this period witnessed the first serious attempts to experimentally verify 
the specific spectroscopic signatures of LLs \cite{Grioni09}. These are the 
(i) low-energy power-law suppression of the local spectral function 
$\rho(\omega) \sim | \omega |^\alpha$ for energies $\omega$ close to the 
chemical potential \cite{Theumann67,Dzyaloshinskii73,Luther74} with 
$\alpha$ depending on the two-particle interaction as well as (ii) the 
appearance of two dispersing features in the momentum resolved spectral 
function $\rho(k,\omega)$ (spin-charge separation) \cite{Meden92,Schoenhammer93a,Voit93} 
instead of a single quasi-particle peak of a Fermi liquid. For finite temperatures
$T$ the suppression of the spectral weight as a function of $\omega$ 
is cut off by $T$ and one finds the scaling behavior 
$\rho \sim  T^\alpha S_\alpha(\omega/T)$ with a 
$\alpha$-dependent scaling function $S_\alpha$ in which the two 
energy scales $\omega$ and $T$ only enter via their ratio \cite{Schoenhammer93b}. 
These results were exclusively obtained using bosonization within the 
Tomonaga-Luttinger (TL) model 
\cite{Voit95,Giamarchi03,Schoenhammer05}. 

Using the modern language of renormalization group (RG) methods the 
(translational invariant) TL model 
is the exactly solvable effective low-energy fixed point model for a large class 
of metallic 1d correlated electron systems---the LLs \cite{Haldane81}. It thus 
plays the same role as the free Fermi gas in Fermi liquid theory. The model
has two strictly linear branches of right- and left-moving fermions and two-particle
scattering is restricted to processes with small momentum transfer $|q| \ll k_{\rm F}$, with
the Fermi momentum $k_{\rm F}$. These processes as well as the kinetic energy can be written 
as quadratic forms of the densities of right- and left-moving fermions which obey bosonic 
commutation relations. In most calculations in addition the momentum dependence 
of the low-momentum scattering processes $g_2$ and $g_4$ are neglected and momentum 
integrals are regularized in the ultraviolet introducing a cutoff 'by hand' 
(for an exception see Ref.~\cite{Schoenhammer93a}). One can extend the resulting
{\em scale-free,} field theoretical model by allowing for additional two-particle scattering 
processes. These turn out to be RG irrelevant in a wide parameter regime 
\cite{Solyom79}. The most important of these processes is the so-called 
$g_{1,\perp}$-process (in the g-ology 
classification \cite{Solyom79}) with momentum transfer $2 k_{\rm F}$ between 
two scattering fermions of opposite spin. 
 
In some of the early experiments on 1d chains these were obviously interrupted by 
local impurities \cite{Grioni09}. A simple model of an {\em inhomogeneous} LL is the open 
boundary analog of the TL model. Interestingly, a LL is very susceptible towards 
single-particle perturbations with momentum transfer $2 k_{\rm F}$ \cite{Luther74} 
and on asymptotically low energy scales even a single weak impurity has the same 
effects on the spectral properties as an open boundary \cite{Kane92}. Triggered by 
this theoretical insight and the early experiments, the spectral properties of the 
open boundary analog of the TL model were studied \cite{Fabrizio95,Eggert96}. 
The local spectral function close to the boundary shows power-law behavior as 
a function of $\omega$ but with an exponent $\alpha_{\rm B}$ different from 
the bulk one $\alpha$.  
As in the translational invariant case in this model only those low-energy 
scattering terms are kept which can be written as quadratic forms in bosonic 
densities. Only recently it was shown that a large class of further two-particle 
processes appearing in a 1d system with an open boundary are indeed 
RG irrelevant \cite{Grap09}. 

The latest scanning tunneling spectroscopy (STS) and photoemission spectroscopy 
(PES) measurements on different classes of 1d metallic systems \cite{Ishii03,Hager05,Wang06,
Jompol09,Blumenstein11} impressively demonstrated that the experiments caught 
up and more refined questions must now be answered by theory. Important ones are: How 
do effects which are not captured by the low-energy fixed point model, such as the 
momentum dependence of the two-particle interaction and the nonlinearity of the 
single-particle dispersion influence the spectral functions? What is the energy scale of a given 
microscopic model on which the low-energy LL physics sets in? How do scaling 
functions for lattice models look like in detail? Here we shed some light on the 
last two questions and briefly comment on the first one. It is widely 
believed that neglecting the momentum dependence of the interaction and regularizing 
momentum integrals in the ultraviolet 'by hand' has no effect on the low-energy 
physics of LL's. This is indeed correct if all energy scales are sent to zero, that 
is for $\rho(\omega)$ and $\rho(\pm k_{\rm F},\omega)$: at small $\omega$ the 
spectral properties are unaffected by the details of the momentum dependence of 
the $g$'s. However, if $\rho(k,\omega)$ as a function of $\omega$ is studied at fixed 
$k \neq \pm k_{\rm F}$, as it is usually done in angular resolved PES, details 
of the momentum dependence of the interaction matter. This was investigated 
in  Ref.~\cite{Meden99}. An overview on the effects of the nonlinearity of 
the single-particle dispersion can be found in the very recent review Ref.~\cite{Imambekov11}. 

This paper is organized as follows. In Sect. \ref{sec:scalfun}  we compute 
the local spectral function of the  translationally invariant {\em and} the 
open boundary continuum TL model using bosonization. We show that both display scaling 
in $\omega/T$ and that the scaling functions have the same analytic form. 
We next compute the spectral function of 
the extended Hubbard model on the lattice close to an open boundary as a function of energy, 
temperature and position in Sect.~\ref{sec:exthubbard}. For this an 
approximate method is used which is based on the functional 
RG approach \cite{Metzner11}.  It is devised for weak to intermediate 
two-particle interactions. In particular, we concentrate on inhomogeneous 
LLs as the boundary exponent $\alpha_{\rm B}$ 
characterizing the spectral function close to an open boundary is 
{\em linear} in the two-particle interaction while the bulk exponent 
is {\em quadratic} (see below). Varying the microscopic
parameters of the extended Hubbard model we can tune the strength of the different
scattering processes and thus study the crossover between nonuniversal behavior 
and the low-energy LL physics. We perform a scaling analysis of the spectral 
function as a function of $\omega$ and $T$ and show that the spectral weight 
close to the boundary follows the bosonization prediction within the universal 
low-energy regime. The position dependence of the spectral function is characterized 
by oscillatory behavior and a power-law envelope function in accordance with the 
result for the TL. Interestingly, we additionally find a phase shift which is 
proportional to the two-particle interaction and not accounted for in the standard 
bosonization procedure. We summarize our results in Sect.~\ref{sec:summary} and briefly 
comment on how spin-orbit interaction and several bands crossing the Fermi surface---both 
being potentially important effects in recent experiments---influence the 
single-particle spectral functions.         

\section{Scaling functions of the Tomonaga-Luttinger model}
\label{sec:scalfun}

In this section we derive closed analytic expressions for the single-particle 
Green function and the related local spectral function of the TL model with 
and without an open boundary at finite $T$. We then closely inspect the scaling 
form of the spectral functions. 

In field theoretical notation (see e.g.~Ref.~\cite{Schuricht11}) the Hamiltonian 
density of the TL model in spin-charge separated form reads 
\begin{eqnarray}
{\mathcal H}(x) = \sum_{\nu=\mathrm{c,s}}{\cal H}_{\nu}(x), \quad
{\mathcal H}_{\nu } = \frac{v_\nu}{16\pi}
\biggl[\frac{1}{K_\nu}\bigr(\partial_x\Phi_\nu\bigr)^2+
K_\nu\bigr(\partial_x\Theta_\nu\bigr)^2\biggr] ,
\label{eq:hamdens}
\end{eqnarray}
with the canonical Bose fields $\Phi_\nu$ and their dual fields $\Theta_\nu$.
Within the TL model the charge and spin velocities $v_\mathrm{c,s}$ as well as 
the LL parameters $K_\mathrm{c,s}$ are free parameters. If the model is used 
to describe the low-energy physics of an  underlying microscopic model 
they become functions of the corresponding model parameters 
and the band filling \cite{Voit95,Giamarchi03,Schoenhammer05}.
For spin-rotational invariant models on which we focus in the present and
the next section $K_{\rm s} =1$. For repulsive interactions $0 < K_{\rm c} < 1$ while 
$K_{\rm c} >1$ in the attractive case.  Here we exclusively consider the former. 
The field operator $\Psi_\sigma(x)$ 
annihilating an electron with spin direction 
$\sigma=\uparrow,\downarrow$  at position $x$ is decomposed into a right- and a 
left-moving part 
\begin{equation}
\label{eq:lowenergy}
\Psi_\sigma(x) = e^{\ii k_\mathrm{F} x} R_\sigma(x)+
e^{-\ii k_\mathrm{F} x} L_\sigma(x). 
\end{equation}
The imaginary time fields $R_\sigma$ and $L_\sigma$ are 
bosonized according to
\begin{eqnarray}
R^\dagger_\sigma(\tau,x)&=&\frac{\eta_\sigma}{\sqrt{2\pi}}\,
\exp\Bigl(\tfrac{\ii}{2}\phi_\mathrm{c}(\tau,x)\Bigr)\,
\exp\Bigl(\tfrac{\ii}{2}f_\sigma\phi_\mathrm{s}(\tau,x)\Bigr),
\label{eq:bosonizationR}\\*
L^\dagger_\sigma(\tau,x)&=&\frac{\eta_\sigma}{\sqrt{2\pi}}\,
\exp\Bigl(-\tfrac{\ii}{2}\bar{\phi}_\mathrm{c}(\tau,x)\Bigr)\,
\exp\Bigl(-\tfrac{\ii}{2}f_\sigma\bar{\phi}_\mathrm{s}(\tau,x)\Bigr),
\label{eq:bosonizationL}
\end{eqnarray}
where the Klein factors $\eta_\sigma$ satisfy anticommutation rules
$\anticomm{\eta_\sigma}{\eta_\sigma'}=2\delta_{\sigma,\sigma'}$ and
$f_\up=1=-f_\dw$. The fields $\phi_\nu$ and $\bar{\phi}_\nu$
are the chiral components of $\Phi_\nu$ and $\Theta_\nu$,
\begin{equation}
\label{eq:gaussfields}
\Phi_\nu=\phi_\nu+\bar{\phi}_\nu,\quad \Theta_\nu=\phi_\nu-\bar{\phi}_\nu,\quad
\nu=\mathrm{c},\mathrm{s}.
\end{equation}

For the translational invariant TL model the Hamiltonian follows by integrating the density 
Eq.~(\ref{eq:hamdens}) over $\mathbb R$. The TL model with an open boundary is obtained 
by integrating the density Eq.~(\ref{eq:hamdens}) over $x \geq 0$ and employing the 
boundary condition $\Psi_\sigma(x=0)=0$ for the fermionic and $\Phi_\mathrm{c,s}(x=0)=0$ 
for the bosonic fields, respectively. 

We are here interested in the imaginary time-ordered single-particle Green function 
\begin{equation}
\label{eq:GF}
G_{\sigma\sigma'}(\tau,x_1,x_2)=
-\left< \mathcal{T}_\tau\,\Psi_\sigma(\tau,x_1)\,
\Psi_{\sigma'}^\dagger(0,x_2) \right>,
\end{equation}
where $\left< \ldots \right>$ denotes the expectation value in the canonical ensemble. 
From the decomposition Eq.~(\ref{eq:lowenergy}) it follows that
\begin{equation}
\label{eq:GFlowenergy}
G_{\sigma\sigma'} =
e^{\ii k_\mathrm{F}(x_1-x_2)}\,G^{RR}_{\sigma\sigma'}
+e^{-\ii k_\mathrm{F}(x_1-x_2)}\,G^{LL}_{\sigma\sigma'}
+e^{\ii k_\mathrm{F}(x_1+x_2)}\,G^{RL}_{\sigma\sigma'}
+e^{-\ii k_\mathrm{F}(x_1+x_2)}\,G^{LR}_{\sigma\sigma'},
\end{equation}
where, for example $G^{RL}_{\sigma\sigma'}=- \left<  \mathcal{T}_\tau\,
R_\sigma(\tau,x_1)\,L^\dagger_{\sigma'}(0,x_2) \right>$.  As we are
aiming at the local spectral function, we eventually set 
$x_1=x_2=x$. For the translational invariant TL model
the left- and right-moving fermion fields are independent 
and thus $G^{RL}_{\sigma\sigma'}
=G^{LR}_{\sigma\sigma'}=0$. In the presence of an open boundary the above mentioned 
boundary conditions imply $R_\sigma(\tau,x) = -L_\sigma(\tau,-x)$. In this 
case and after setting $x_1=x_2=x$ the cross terms are 
characterized by a fast spatial oscillation with frequency $2 k_{\rm F}$. 

Using standard methods (see e.g.~Ref.~\cite{Giamarchi03}) one obtains for the Green 
function of a translational invariant system
($K_{\rm s}=1$, $\tau>0$, $\beta=1/T$, $r=x_1-x_2$, $R= (x_1 +x_2)/2$)
\begin{eqnarray}
\mbox{} \hspace{-1.5cm} 
G^{RR}_{\sigma\sigma'}(\tau,x_1,x_2) & = & - \frac{\delta_{\sigma,\sigma'}}{2 \pi} 
\left(\frac{\pi}{v_{\rm c} \beta } \right)^{a+b} 
\left(\frac{\pi}{v_{\rm s} \beta } \right)^{1/2}
\frac{1}{\sin^{1/2}\left[ \frac{\pi}{v_{\rm s} \beta} \left( v_{\rm s} \tau - \ii r \right) \right]} \\
&& \times  
\frac{1}{\sin^a\left[ \frac{\pi}{v_{\rm c} \beta} \left( v_{\rm c} \tau - \ii r \right) \right]} \,
\frac{1}{\sin^b\left[ \frac{\pi}{v_{\rm c} \beta} \left( v_{\rm c} \tau + \ii r \right) \right]} \, 
\nonumber \\
\mbox{} \hspace{-1.5cm} 
G^{LL}_{\sigma\sigma'}(\tau,x_1,x_2) & = &  G^{RR}_{\sigma\sigma'}(\tau,x_2,x_1) 
\end{eqnarray}
and for the case with boundary
\begin{eqnarray}
\mbox{} \hspace{-2.5cm} 
\left[G^{RR}_{\sigma\sigma'}\right]_{\rm B}(\tau,x_1,x_2) & = & 
G^{RR}_{\sigma\sigma'} (\tau,x_1,x_2) \, 
\left\{ \frac{\sinh\left( \frac{2 \pi}{v_{\rm c} \beta} x_1 \right)
\sinh\left( \frac{2 \pi}{v_{\rm c} \beta} x_2 \right)  }
{\sin\left[ \frac{\pi}{v_{\rm c} \beta} \left( v_{\rm c} \tau - 2 \ii R \right) \right]
\sin\left[ \frac{\pi}{v_{\rm c} \beta} \left( v_{\rm c} \tau + 2 \ii R \right) \right]}  \right\}^c \\
\mbox{} \hspace{-2.5cm} 
\left[G^{LL}_{\sigma\sigma'}\right]_{\rm B}(\tau,x_1,x_2) & = &  
\left[ G^{RR}_{\sigma\sigma'}\right]_{\rm B}(\tau,x_2,x_1) 
\end{eqnarray}
The cross terms $\left[G^{RL}_{\sigma\sigma'}\right]_{\rm B}$ and 
$\left[G^{LR}_{\sigma\sigma'}\right]_{\rm B}$ are equal to  $-\left[G^{RR}_{\sigma\sigma'}\right]_{\rm B}$ 
and $-\left[G^{LL}_{\sigma\sigma'}\right]_{\rm B}$ after interchanging $r \leftrightarrow 2 R$. 
The appearing exponents are given by 
\begin{eqnarray}
\mbox{} \hspace{-1.5cm}
a= \frac{1}{8} \left( \sqrt{K_{\rm c}} + \frac{1}{\sqrt{K_{\rm c}}} \right)^2 , \quad
b= \frac{1}{8} \left( \sqrt{K_{\rm c}} - \frac{1}{\sqrt{K_{\rm c}}} \right)^2 , \quad
c= \frac{1}{8} \left( \frac{1}{K_{\rm c}} - K_{\rm c} \right) . 
\label{eq:expcdef} 
\end{eqnarray} 
  
The main steps to obtain the local spectral function from the imaginary time Green function 
are the analytic continuation $\tau \rightarrow \ii t + \delta$ followed by Fourier 
transformation with respect to $t$.  Mathematically the real part $\delta$ corresponds to the 
ultraviolet cutoff introduced to regularize momentum intergrals. From an experimental perspective 
it can be considered as the resolution of the setup (at $T=0$). 

For the translational invariant 
model one obtains ($x_1=x_2=x$)
\begin{eqnarray}
\mbox{} \hspace{-2.cm}
\rho(\omega,x) & = &  - \frac{1}{2 \pi} \left( 1+ e^{-\beta \omega} \right) \sum_{\sigma,\sigma'} 
 \left. \int_{-\infty}^\infty dt \, e^{\ii \omega t} \left[ G^{RR}_{\sigma\sigma'}(\tau,x,x)
+ G^{LL}_{\sigma\sigma'}(\tau,x,x) \right] \right|_{\tau \rightarrow \ii t+ \delta}  \nonumber \\
\mbox{} \hspace{-2.cm}
& = &  \frac{1}{\pi^2}   \left( 1+ e^{-\beta \omega} \right) \left( \frac{\pi}{v_{\rm c} \beta} \right)^{a+b} 
 \left( \frac{\pi}{v_{\rm s} \beta} \right)^{1/2}  
 \left. \int_{-\infty}^\infty dt \, \frac{ e^{\ii \omega t}}{\sin^{a+b+1/2} \left( \frac{\pi \tau}{\beta} \right)  }
 \right|_{\tau \rightarrow \ii t+ \delta} \nonumber \\
& = & \frac{4 \pi^{a+b+1/2}}{\pi^2 v_{\rm c}^{a+b} v_{\rm s}^{1/2}} \, T^\alpha \, S_\alpha(\omega/T) ,
\label{eq:rhotransinv}
\end{eqnarray}
with 
\begin{eqnarray}
&& S_\gamma(u) =  2^{1+ \gamma} \Gamma(-\gamma) \sin{[\pi(1+\gamma)]} 
\cosh{\left(  \frac{u}{2} \right)} \, \left| \Gamma \left( \frac{1+\gamma}{2}+\ii \frac{u}{2 \pi}\right) \right|^2 , 
\label{eq:Sdef} \\
&& \alpha  =  a+b-\frac{1}{2} = \frac{1}{4} \left( K_{\rm c} + \frac{1}{K_{\rm c}} -2 \right) . 
\label{eq:diveresedeffs} 
\end{eqnarray}
The {\em position independent} local spectral weight of the translational invariant TL model thus shows 
{\em scaling behavior}: $T^{- \alpha} \, \rho(\omega,x)$ is a function of $\omega/T$ only. The amplitude 
of the scaling function depends on $v_{{\rm c}/{\rm s}}$ and $K_{\rm c}$, which in turn are 
functions of the interaction strength, while its shape is given by 
$K_{\rm c}$ only. This result was first derived in Ref.~\cite{Schoenhammer93b}. A similar 
expression was later used to describe transport properties of LLs \cite{Bockrath99}. 

Taking the $T \to 0$ limit of Eq.~(\ref{eq:rhotransinv}) one obtains the well known 
power-law suppression of the spectral weight \cite{Theumann67,Dzyaloshinskii73,Luther74}
\begin{eqnarray}
T=0: \quad \rho(\omega,x) \sim |\omega|^\alpha , 
\end{eqnarray} 
for $|\omega| \to 0$. For fixed small $T>0$ this is cut off by temperature and $\rho$ saturates 
for $|\omega | \lessapprox T$. For the energy set to the chemical potential, that is 
$\omega=0$, one finds a power-law suppression of $\rho$ for $T \to 0$ 
(see Eq.~(\ref{eq:rhotransinv})): $\rho \sim T^{\alpha}$.   
From studies of microscopic models it is known that (see e.g.~Ref.~\cite{Schoenhammer05})
\begin{eqnarray}
K_{\rm c} = 1 - \frac{U}{U_{\rm m}} + {\mathcal O}\left( \left[\frac{U}{U_{\rm m}}\right]^2\right),
\label{eq:Kexp}
\end{eqnarray}
where $U$ is a measure of the two-particle interaction and $U_{\rm m}$ is a scale which depends on the 
other model parameters. Using Eq.~(\ref{eq:diveresedeffs}) one thus obtains
\begin{eqnarray}
\alpha \sim U^2 ,
\label{eq:U_2} 
\end{eqnarray}
for the exponent characterizing the low-energy behavior of the local spectral function 
of the translational invariant TL model.

Verifying the power-law suppression of the spectral weight as a function 
of $T$ and $\omega$ with the same exponent $\alpha$ as well as the  scaling property 
of measured STS and/or PES data provide strong indications that the system under 
investigation is indeed a LL \cite{Blumenstein11}. It was very recently argued that 
these characteristics are still not unique to LLs as other mechanisms than 1d electronic 
correlations might lead to similar behavior \cite{Rodin10}. We therefore suggest further 
consistency checks by in addition measuring spectra close to the end points of cut 1d chains.

The local spectral function becomes position $x$ dependent when considering a chain with an 
open boundary. In this case $\rho(\omega,x)$ has three contributions
\begin{eqnarray}
\rho_{\rm B}(\omega,x) = \rho_0(\omega,x) + e^{2 \ii k_{\rm F} x}  \rho_{2 k_{\rm F}}(\omega,x) 
+ e^{-2 \ii k_{\rm F} x}  \rho_{-2 k_{\rm F}}(\omega,x) ,
\label{eq:summe}
\end{eqnarray} 
where the first follows from Fourier transforming $ G^{RR}+ G^{LL}$ and the last two from 
transforming $ G^{RL}$ and  $ G^{LR}$, respectively ($x_1=x_2=x$). 
Following the same steps as above 
we obtain
\begin{eqnarray}
\rho_0(\omega,x) =   \frac{\pi^{a+b+1/2}}{\pi^2 v_{\rm c}^{a+b} 
v_{\rm s}^{1/2}} \, T^{a+b-1/2} \, F(\omega/T,xT/v_{\rm c}) ,
\label{eq:term1}
\end{eqnarray} 
with
\begin{eqnarray}
\mbox{} \hspace{-2.cm}
F(u,v) = \int_{-\infty}^{\infty}    ds \frac{e^{\ii u s} \left(1+e^{-u}  \right)}{\sin^{a+b+1/2}
(\ii \pi s + \delta)} \left( \frac{\sinh^2(2 \pi v)}{\sin(\ii \pi s - 2 \pi \ii v +\delta)
\sin(\ii \pi s + 2 \pi \ii v +\delta)} \right)^c
\end{eqnarray} 
and 
\begin{eqnarray}
\rho_{2 k_{\rm F}}(\omega,x) = \rho_{-2 k_{\rm F}}^\ast(\omega,x) =   
-\frac{\pi^{a+b+1/2}}{2 \pi^2 v_{\rm c}^{a+b} 
v_{\rm s}^{1/2}} \, T^{a+b-1/2} \, G(\omega/T,xT/v_{\rm c}) ,
\label{eq:term2}
\end{eqnarray} 
where
\begin{eqnarray}
\mbox{} \hspace{-2.5cm}
G(u,v) = \int_{-\infty}^{\infty}    ds \,  &&  \frac{e^{\ii u s} \left(1+e^{-u}  \right)}{\sin^{1/2}
(\ii \pi s - 2 \pi \ii v v_{\rm c}/v_{\rm s}+ \delta)} \,
\frac{1}{{\sin^{a}(\ii \pi s - 2 \pi \ii v + \delta)}} \,
\frac{1}{{\sin^{b}(\ii \pi s + 2 \pi \ii v + \delta)}}  \nonumber \\
&& \times \left( \frac{\sinh(2 \pi v)}{\sin(\ii \pi s +\delta)} \right)^{2c} .
\label{eq:Gdef}
\end{eqnarray} 
In the limit $T \to 0$ these expressions simplify to the ones given in 
Refs.~\cite{Eggert96} and \cite{Eggert00}.

For distances from the boundary beyond the thermal length $\sim 1/T$, that is 
$xT/v_{\rm c} = v \gg 1$ we expect $\rho_{\rm B}(\omega,x)$ to become equal to 
$\rho(\omega,x)$ of Eq.~(\ref{eq:rhotransinv}) (exponentially fast). That this is 
indeed the case follows from
\begin{eqnarray}
\frac{x T}{v_{\rm c}}\gg 1: \quad  F(\omega/T,xT/v_{\rm c}) \approx 4 S_{\alpha}(\omega/T) , \quad 
 G(\omega/T,xT/v_{\rm c}) \approx 0 , 
\label{eq:indeed}
\end{eqnarray}
Eqs.~(\ref{eq:summe}), (\ref{eq:term1}), (\ref{eq:term2}) and (\ref{eq:Sdef}). We thus end up with
\begin{eqnarray}
\frac{x T}{v_{\rm c}}\gg 1: \quad \rho_{\rm B} (\omega,x) \approx  \rho (\omega,x) .
\end{eqnarray}

We next consider the limit $xT/v_{\rm c} = v \ll 1$, that is the local spectral 
function close to the open boundary. Then
\begin{eqnarray}
F(u,v) \approx  (2 \pi)^{2c} v^{2c} \left(1+e^{-u}  \right)
\int_{-\infty}^{\infty}    ds \frac{e^{\ii u s}}{\sin^{a+b+2c+1/2}
(\ii \pi s + \delta)} .
\end{eqnarray} 
Interestingly the remaining integral has the same form as the one appearing in the second line 
of  Eq.~(\ref{eq:rhotransinv}) but with $a+b+1/2$ replaced by $a+b+2c+1/2$. For 
fixed $x$ close to the boundary $\rho_0(x,\omega)$ thus displays scaling with the {\em same} scaling 
function as the one found in the bulk but $\alpha$ replaced by 
\begin{eqnarray}
\alpha_{\rm B} = a+b+2c-1/2 = \frac{1}{2} \left( \frac{1}{K_{\rm c}}-1 \right) .
\label{eq:alphaBdef}
\end{eqnarray}
Explicitely one obtains 
\begin{eqnarray} 
\mbox{} \hspace{-1.5cm}
\frac{x T}{v_{\rm c}}\ll 1:  \quad \rho_0(x,\omega) \sim  x^{2c} \, T^{\alpha_{\rm B}} S_{\alpha_{\rm B}}(\omega/T) . 
\end{eqnarray}
With Eq.~(\ref{eq:Kexp}) the boundary exponent $\alpha_{\rm B}$ Eq.~(\ref{eq:alphaBdef}) has, in 
contrast to the bulk one $\alpha$, a contribution {\em linear} in the interaction
\begin{eqnarray}
\alpha_{\rm B} \sim U  
\label{eq:U_1}
\end{eqnarray} 
and one finds $\alpha_{\rm B} > \alpha$. 
To show that for fixed $x$ close to the boundary $T^{-\alpha_{\rm B}} \rho_{\rm B}$ indeed 
follows the same scaling function as in the bulk (with $\alpha \to \alpha_B$) we still have 
to analyze $\rho_{2 k_{\rm F}}$ for $xT/v_{\rm c} = v \ll 1$. In the simplest approximation we 
neglect the $v$-dependence in the integral Eq.~(\ref{eq:Gdef}) and obtain
\begin{eqnarray}
\frac{x T}{v_{\rm c}}\ll 1:  \quad \rho_{2 k_{\rm F}}(\omega,x) = - \, \frac{1}{2}  \rho_{0}(\omega,x) .
\end{eqnarray}  
Using Eq.~(\ref{eq:summe}) this completes our proof that for fixed
\begin{eqnarray}
x \; \mbox{in the bulk:} && \quad T^{- \alpha} \rho(\omega,x) =   T^{- \alpha} \rho_{\rm B}(\omega,x) 
\sim S_\alpha(\omega/T), \\
x \; \mbox{close to the boundary:} && \quad T^{-\alpha_{\rm B}} \rho_{\rm B}(\omega,x)   
\sim S_{\alpha_{\rm B}}(\omega/T) ,
\label{eq:Tgives}
\end{eqnarray}
with $S_\gamma$ given in Eq.~(\ref{eq:Sdef}).

Showing the consistency of the scaling of spectra measured in the two spatial regimes is within 
reach of the latest STS experiments \cite{Blumenstein11}. Combined with a consistency check of 
the two exponents $\alpha$ and $\alpha_{\rm B}$, which both depend on $K_{\rm c}$ only and which 
was already achieved in Ref.~\cite{Blumenstein11} (see also Sect.~\ref{sec:summary}), this would 
provide a stringent experimental verification of LL physics. 

One can {\em improve} the analysis of $\rho_{2 k_{\rm F}}$ close to the boundary by keeping the 
phase factor $\exp(\ii \kappa u v)$ of the integral 
Eq.~(\ref{eq:Gdef}).\footnote{We have verified numerically that the additional dependences of the
integral Eq.~(\ref{eq:Gdef}) on $u$, $v$, and the model parameters are irrelevant 
for the regimes studied here.} A numerical 
evaluation of the integral shows that $\kappa$ is a function of $K_{\rm c}$ and 
$v_{\rm c}/v_{\rm s}$ with $\kappa(K_{\rm c}=1,v_{\rm c}/v_{\rm s}=1)=2$. Taking all terms 
together we find
\begin{eqnarray}
\mbox{} \hspace{-1.5cm}
\frac{x T}{v_{\rm c}}\ll 1:  \quad T^{-\alpha_{\rm B}} \, \rho_{\rm B}(\omega,x) = && A \, x^{2c} \, 
\cosh{ \left( \frac{\omega}{2T} \right)} \, \left| \Gamma \left( \frac{1+\alpha_{\rm B}}{2}+\ii \frac{\omega}{2 \pi T}\right) \right|^2
\nonumber \\ 
&& \times \left\{ 1- \cos\left[ \left( 2 k_{\rm F} + \frac{\kappa \omega}{v_{\rm c}} \right) 
x \right] \right\},
\label{eq:rhocompl}
\end{eqnarray} 
where the overall amplitude $A$ depends on $K_{\rm c}$, $v_{\rm c}$ and $v_{\rm s}$ but {\em not} on the 
variables $\omega$, $x$, and $T$. 

In the $T \to 0$ limit Eqs.~(\ref{eq:summe}) to (\ref{eq:Gdef}) give close to 
the boundary
\begin{eqnarray}
\mbox{} \hspace{-1.5cm}
T=0, \; \frac{x |\omega|}{v_{\rm c}}\ll 1:  
\quad \rho_{\rm B}(\omega,x) \sim  x^{2c} \, |\omega|^{\alpha_{\rm B}} 
\left\{ 1- \cos\left[ \left( 2 k_{\rm F} + \frac{\kappa \omega}{v_{\rm c}} \right) x \right] \right\} .
\label{eq:rhocomplT0}
\end{eqnarray} 
At fixed $x$, $ \rho_{\rm B}$ thus vanishes $\sim  |\omega|^{\alpha_{\rm B}} $. 
As in the translational invariant case this power law is cut off by a finite 
temperature and $ \rho_{\rm B}$ saturates for $|\omega| \lessapprox T$. For 
fixed $x$ close to the boundary and $\omega=0$ Eq.~(\ref{eq:Tgives}) gives 
$\rho_{\rm B} \sim T^{\alpha_{\rm B}}$ for $T \to 0$. 
At $T=0$ and deep in the bulk, that is for $x |\omega| / v_{\rm c} \gg 1$,  
$\rho_{\pm 2 k_{\rm F}}$ can be written as a sum of terms which vanish algebraically 
in $x$ \cite{Eggert96,Eggert00}. They show a power-law dependence on 
$|\omega|$ in general each with a different exponent. The contribution
$\rho_0$ to $\rho_{\rm B}$ becomes position independent and goes 
as $\rho_0 \sim |\omega|^{\alpha}$ (instead of $\alpha_{\rm B}$ close to 
the boundary). For sufficiently large $x |\omega| /v_{\rm c}$ 
(such that algebraically decaying terms can be neglected) one thus finds ($\omega \to 0$)   
\begin{eqnarray}
\mbox{} \hspace{-1.5cm}
T=0, \; \frac{x |\omega|}{v_{\rm c}}\gg 1:  \quad \rho_{\rm B}(\omega,x) \sim  |\omega|^{\alpha} . 
\label{eq:rhocomplT0_bulk}
\end{eqnarray} 
For $T=0$ and in the {\em noninteracting} limit $\rho^0_{\pm 2 k_{\rm F}}$ does not decay 
and one obtains 
\begin{eqnarray}
T=0: \quad \rho_{\rm B}^0(\omega,x) = \frac{2}{\pi v_{\rm F}} \,  
\left\{ 1- \cos\left[ 2 \left( k_{\rm F} + \frac{\omega}{v_{\rm F}} \right) x \right] \right\} ,
\label{eq:rhocompl0}
\end{eqnarray} 
for all $x$ and $\omega$.

It is often argued, that the contribution $\rho_{2 k_{\rm F}}$ to $\rho_{\rm B}$ can be neglected when 
comparing to experiments. The electrons in PES and STS do not come from a specific location $x$ but rather 
from an extended spatial range. If this is large enough compared to the characteristic length 
$1/k_{\rm F}$, $\rho_{2 k_{\rm F}}$ averages out due to the fast spatial oscillations with frequency 
$2 k_{\rm F}$. It is not obvious that the criterion 
for neglecting $\rho_{2 k_{\rm F}}$ is fulfilled in the latest STS experiments \cite{Blumenstein11} 
(see Sect.~\ref{sec:summary}).  

Equation (\ref{eq:rhocompl}) (or Eq.~(\ref{eq:rhocomplT0}) for $T=0$) allows for another consistency 
check of LL behavior. It predicts a spatial power-law dependence of $\rho_{\rm B}$ 
close to the boundary with exponent $2c = (1/K_{\rm c} - K_{\rm c})/4$  
superimposed by oscillations. If it is possible to measure the envelope function of the 
spatial dependence of the spectral weight close to a boundary and extract the power-law exponent 
it would allow to relate the resulting $K_{\rm c}$ to the ones obtained from $\alpha$ 
and/or $\alpha_{\rm B}$. 

We note in passing that performing a spatial Fourier transform of $\rho_{\rm B}(\omega,x)$ reveals
characteristic informations of the bulk state of a LL including its elementary excitations 
(see Ref.~\cite{Schuricht11} and references therein). 
 
In the next section we show that the spectral function of microscopic lattice models of interacting 1d 
electrons, in our case the extended Hubbard model, indeed shows scaling behavior as a function
of $\omega/T$. Up to a subtlety in the spatial dependence, namely an interaction dependent 
phase shift in the oscillatory factor, the lattice spectral function 
falls on top of the above computed scaling function of the TL model. This holds in the low-energy 
regime. We discuss the crossover between this universal behavior and the nonuniversal regime 
at higher energies. The crossover scale $\Delta$ depends on the parameters of the microscopic model.    

\section{Spectral properties of the extended Hubbard model}
\label{sec:exthubbard}

Partly aiming at the low-energy LL physics of microscopic lattice models of interacting 
electrons different groups computed the local spectral function 
\cite{Preus94,Benthien05} as well as the momentum resolved one \cite{Benthien04,Abendschein06} 
for such models using very accurate numerical methods (often denoted as 'numerically exact'). 
Quantum Monte-Carlo (QMC) \cite{Preus94,Abendschein06} and (dynamical) density-matrix renormalization group (DMRG) 
\cite{Benthien04,Benthien05} were used. The results obtained 
by these methods turned out to be very useful for understanding the spectral features of the studied 
models over the entire band width. Unfortunately, due to system size restrictions (DMRG and QMC), 
artificial broadenings of the spectra (DMRG), as well as the problem of analytic continuation of 
numerical data (QMC), it was not possible to reach the low-energy regime; in none of the calculations 
it was possible to convincingly demonstrate power-law behavior of the spectra. Rephrasing this in 
experimental terms one can say that the energy resolution of these methods is not high 
enough.\footnote{For a very recent attempt to observe power-law behavior of the spectral function 
in a spinless lattice model using DMRG, see Ref.~\cite{Jeckelmann11}.}
For (partly) technical reasons the focus of the numerical approaches lies on the Hubbard model with 
a local two-particle interaction. As the crossover scale $\Delta$ between LL behavior and 
nonuniversal physics in this model is very small (see below) reaching the LL regime 
is particularly challenging.      

We here use a method which allows to obtain {\em approximate} results for the spectral function 
of the extended Hubbard model with a local $U$ and nearest-neighbor $V$ interaction. Our approximation 
is based on the functional RG approach to quantum many-body physics  \cite{Metzner11}. Functional RG 
allows to set up a hierarchy of approximation schemes with the two-particle interaction being the small 
parameter. The one we are using here is controlled to leading order and can thus only be used 
for small to intermediate $U$ and $V$ (compared to the band width). The method 
was mainly applied in the context of transport 
through inhomogeneous LLs and there it was shown to reproduce typical impurity strength 
independent \cite{Kane92} LL exponents to {\it leading order} in the interaction 
\cite{Enss05,Andergassen06}. Due to appropriate resummations of classes 
of diagrams the RG procedure thus goes way beyond standard perturbation theory. 
As the exponent of the bulk local spectral function is of {\em second order} in the 
interaction (see Eq.~(\ref{eq:U_2})) the LL physics of 
translationally invariant systems cannot be assessed in our approximation. In the 
following we therefore restrict ourselves to the model with open boundaries characterized by the 
exponent $\alpha_{\rm B}$ which has a {\em linear} contribution (see Eq.~(\ref{eq:U_1})) and study 
$\rho_{\rm B}$ {\em close} to one of the boundaries. We here refrain from giving any 
further technical details 
on how the spectral function can be computed within our functional RG approach. Those can be found in 
Ref.~\cite{Andergassen06}. 
      
The Hamiltonian of the extended Hubbard model with two open boundaries is given by 
\begin{equation}
 H = -t \sum_{j=1}^{N-1} \sum_{\sigma} \big( \,
 c^{\dag}_{j+1,\sigma} c_{j,\sigma}^{\phantom{\dag}} + 
 c^{\dag}_{j,\sigma} c_{j+1,\sigma}^{\phantom{\dag}} \, \big) \; + 
 U \sum_{j=1}^{N} n_{j,\uparrow} \, n_{j,\downarrow} + V \sum_{j=1}^{N-1} n_j \, n_{j+1} ,
\label{eq:model}
\end{equation}
where $c^{\dag}_{j,\sigma}$ and $c_{j,\sigma}^{\phantom{\dag}}$ are 
creation and annihilation operators for fermions with spin 
$\sigma$ on lattice site $j$, while
$n_{j,\sigma} = c^{\dag}_{j,\sigma} \, c_{j,\sigma}^{\phantom{\dag}} \,$,
and $n_j = n_{j,\uparrow} + n_{j,\downarrow}$ is the local density operator 
on site $j$. For the (nonextended) Hubbard model the nearest neighbor 
interaction $V$ vanishes. The number of lattice sites is denoted by $N$. 
The noninteracting tight-binding part gives the standard  
dispersion $\epsilon_k = -2t \cos{ k }$ with the hopping matrix element $t>0$ 
(the lattice constant is chosen to be unity). 

Under the {\em assumption} that a given microscopic model is a LL (at low 
energy scales) one can use general relations between the exact ground state 
energy $E_0$ and $K_{\rm c}$ \cite{Haldane81,Voit95,Giamarchi03,Schoenhammer05} to  
extract the dependence of the LL parameter $K_{\rm c}$ on the parameters of the model
considered. In general however $E_0$ of a many-body system is not known 
analytically. The translational invariant Hubbard model constitutes one of the
rare exceptions and closed expressions for $E_0$ in form of integral equations 
can be determined using Bethe ansatz \cite{Essler05}. The integral equations can easily be solved 
numerically which gives access to the dependence of $K_{\rm c}$ on the parameters 
$U/t$ and the band filling $n$ ($n$ can vary between 0 and 2) \cite{Schulz90}. We 
emphasize that this {\em only} implies that on {\em asymptotically small} scales one 
can expect power-law behavior with $\alpha$ (bulk) or 
$\alpha_B$ (close to the boundary) while no information 
on the crossover scale $\Delta$ from nonuniversal to universal LL behavior can 
be extracted this way. Furthermore, this expectation holds under the assumption 
that the Hubbard model is a LL, which away from half-filling $n=1$---for which it 
is a Mott insulator \cite{Essler05}---is not doubted seriously, but also not proven rigorously. 
For the (translational invariant) extended Hubbard model $K_{\rm c}(U/t,V/t,n)$ 
can only be computed 
numerically along similar lines \cite{Ejima05}. The exponents $\alpha$ and $\alpha_{\rm B}$ 
for the Hubbard \cite{Schulz90} cannot become large enough to match the exponents inferred 
from experiments on different systems \cite{Ishii03,Hager05,Wang06,Jompol09,Blumenstein11}, 
or, putting it differently, $K_{\rm c}$ cannot become small enough. For the extended 
Hubbard model $K_{\rm c}$'s of roughly the correct order can be achieved 
for $U$ and $V$ of the order of the band width or larger. This part of the parameter 
space lies very close to the Mott transition of the model \cite{Ejima05}. 
One can expect that this effects the spectral properties. When aiming at a typical LL 
spectral function with $\alpha$ and $\alpha_{\rm B}$ of experimental size it is thus 
advisable to study models with interaction of longer spatial range. We note that within 
our approximate approach $U$ and $V$ are bound to be sufficiently smaller than the 
band width.  The extended Hubbard model is 
spin-rotational invariant which implies $K_{\rm s}=1$.

\begin{figure}[t]
\begin{center}
   \includegraphics[width=0.7\linewidth,clip]{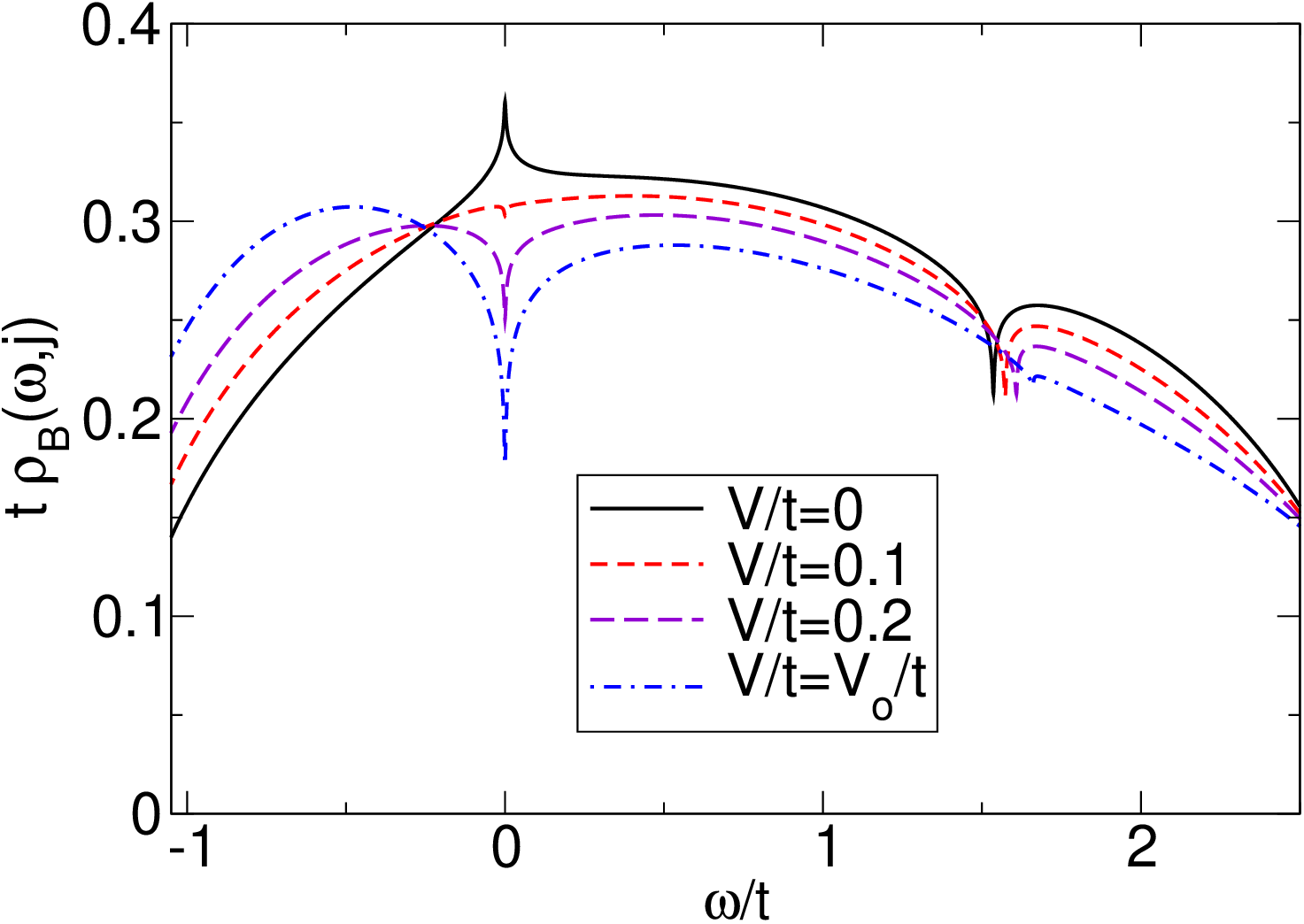}
   \caption{(Color online) Local spectral function of the extended Hubbard model for 
    $n=3/4$, $j=1$, $U/t=0.5$, $N=2^{14}$, $T/t=10^{-3}$ and different $V/t$. For filling 
    $n=3/4$ the optimal nearest-neighbor interaction is given by 
    $V_{\rm o}/t=U/(t \sqrt{2})\approx 0.35$. Only for sizable nearest-neighbour 
    interaction $V$ we observe the LL suppression of the weight at $\omega=0$. The suppression 
    close to $\omega/t =1.5$ is a lattice effect \cite{Andergassen04}.}
   \label{fig1}
\end{center}
\end{figure}

For the Hubbard model with an open boundary the scale $\Delta$ was earlier computed 
in the small $U$ limit and it was estimated to be exponentially small \cite{Meden00}
\begin{eqnarray}
\frac{\Delta}{v_{\rm F} k_{\rm F}} = \exp{\left\{ -  \frac{\pi v_{\rm F}}{U} 
\ln{\frac{1+ [U/(8 v_{\rm F})]^2}{ [U/(8 v_{\rm F})]^2}} \right\}} .
\label{eq:crossoverscale}
\end{eqnarray}
In fact, approaching 
the chemical potential $\omega \to 0$ the local spectral weight first increases 
before the LL power-law suppression sets in for $|\omega | \lessapprox \Delta$
(see the solid line in Fig.~\ref{fig1} which does not look 'LL-like' around $\omega=0$; 
the power-law suppression is beyond the energy resolution).   
This is consistent with the observation of a  small crossover scale
and a peak close to $\omega=0$ in the local spectra of the 
{\em translational invariant} Hubbard model 
obtained numerically by QMC and DMRG \cite{Preus94,Benthien05}. 
   
To compute the finite temperature $\rho_{\rm B}(\omega,j)$ (here the continuous position 
$x$ is replaced by the discrete lattice site index $j$) of the extended Hubbard model 
we consider a chain of $N$ lattice sites described by Eq.~(\ref{eq:model}).
For this the spectrum is discrete and the spectral function consists of 
$\delta$-peaks. Due to even-odd effects the spectral weight might vary quickly 
from one eigenvalue to the next one. A smooth function of $\omega$ is obtained by
averaging the weight over neighboring eigenvalues.\footnote{A similar energy averaging 
is inherent to any STS or PES experiment due to the finite energy resolution 
of these techniques.} To obtain the local spectral 
function as defined in the continuum one furthermore has to devide the weights 
by the level spacing between eigenvalues. The energy scale $\delta_N = \pi v_{\rm F}/N$ 
associated to the chain length becomes irrelevant as we always consider sufficiently 
large systems with $T \gtrapprox \delta_N$ for fixed $T$. Our results are thus not 
influenced by finite size effects (for an exception, see the discussion of Fig.~\ref{fig4}). 
Typical experimental temperatures are in the 
few to few ten Kelvin range which corresponds to $T \approx 10^{-4} t$ to $10^{-3} t$ 
for our model. 
    
In Fig.~\ref{fig1} $\rho_{\rm B}(\omega,j)$ is shown for filling $n=3/4$, 
lattice site $j=1$ next to the boundary, $U/t=0.5$, $N=2^{14}$, $T/t=10^{-3}$ and 
different $V/t$. 
For the Hubbard model with $V/t=0$ no suppression of the spectral 
weight is observable as $\Delta$ is much smaller than 
temperature. Obviously, $\Delta$ increases with increasing $V/t$ and 
the LL suppression at $\omega=0$ becomes apparent. This can be understood as 
follows. The crossover scale $\Delta$ is strongly affected by the size of the open boundary analog 
of a $g_{1,\perp}$ two-particle scattering process \cite{Grap09,Andergassen06,Meden00} which 
cannot be written quadratically in the bosonic densities. Its initial value (with respect to an 
RG flow) in the extended Hubbard model is given by $g_{1,\perp} = U + 2 V \cos(2 k_{\rm F})$ 
with $k_{\rm F} = n \pi/2$. 
It is large for $V=0$. At $V_{\rm o}=-U/[2 \cos(2 k_{\rm F})]$ it vanishes (see the dashed-dotted line 
in Fig.~\ref{fig1}). Under an RG procedure this 'non-LL term' flows to zero and is thus RG 
irrelevant. However, the flow is only logarithmically. This implies that for sizable 
initial $g_{1,\perp}$ LL physics sets in on exponentially small scales consistent with 
Eq.~(\ref{eq:crossoverscale}) for the Hubbard model \cite{Meden00}. For small initial 
$g_{1,\perp}$, $\rho_{\rm B}$ appears LL-like with the characteristic power-law behavior of the spectral 
weight close to $\omega=0$. In Fig.~\ref{fig1} the spectral weight at $\omega=0$ remains finite due to 
the finite temperature. We conclude that to observe LL physics on moderate scales in the present model 
the interaction should not be too local. In particular, to demonstrate power-law behavior and obtain 
an estimate of the exponent by fitting $\rho_{\rm B}$ for fixed $j$ (close to the boundary) 
and $T \ll t$ as a function 
of $\omega$ (in the range $T \ll |\omega| \ll t$) one should consider fine-tuned 
parameters with $V=V_{\rm o}$.
The spectral functions of Fig.~\ref{fig1} show another 'high-energy' 
nonanalyticity. A similar feature was observed for a model of spinless fermions in 
Ref.~\cite{Andergassen04} and was explained there as a lattice effect.

\begin{figure}[t]
\begin{center}
   \includegraphics[width=0.7\linewidth,clip]{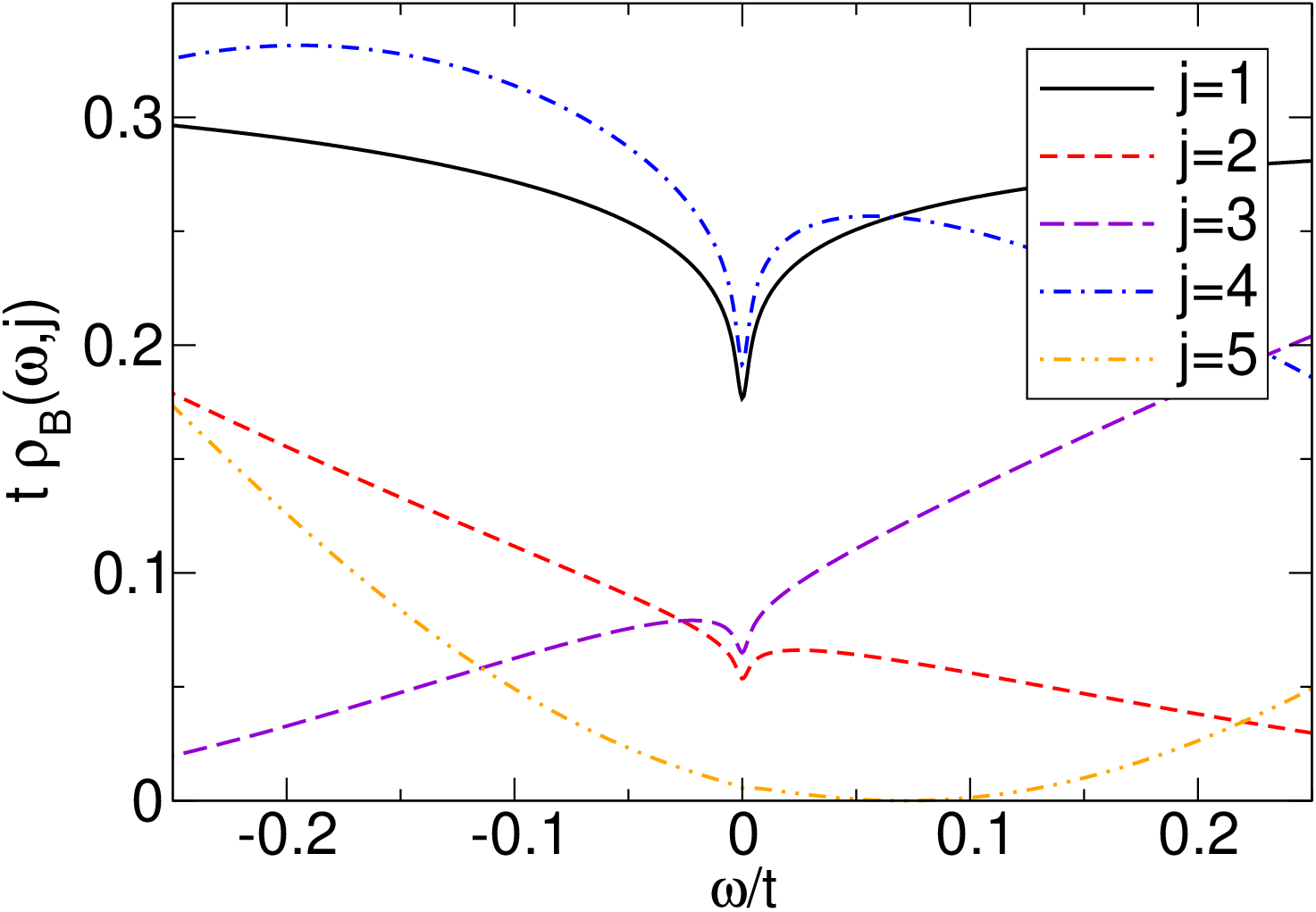}
   \caption{(Color online) Low-energy regime of the local spectral function 
     of the extended Hubbard model for  $n=3/4$, $U/t=0.5$, $V=V_{\rm o} = U/\sqrt{2}$ 
     (for $n=3/4$), 
     $N=2^{14}$, $T/t=10^{-3}$ and  different $j$ close 
     to the open boundary at $j=1$.}
   \label{fig2}
\end{center}
\end{figure}

In Fig.~\ref{fig2} the low-energy regime of  $\rho_{\rm B}$ is shown for the same parameters 
as in Fig.~\ref{fig1} but  'optimal' $V=V_{\rm o}=U/\sqrt{2}$ (for $n=3/4$), which 
allows for the largest low-energy regime, and varying position $j$ close to the 
boundary site $j=1$. For fixed $\omega$ we observe 
strong variations of the weight with $j$ and pronounced $\omega \leftrightarrow - \omega$ asymmetries 
which is consistent with the result from the TL model Eq.~(\ref{eq:rhocompl}). Below we return 
to the spatial dependence of $\rho_{\rm B}$. 

\begin{figure}[t]
\begin{center}
   \includegraphics[width=0.7\linewidth,clip]{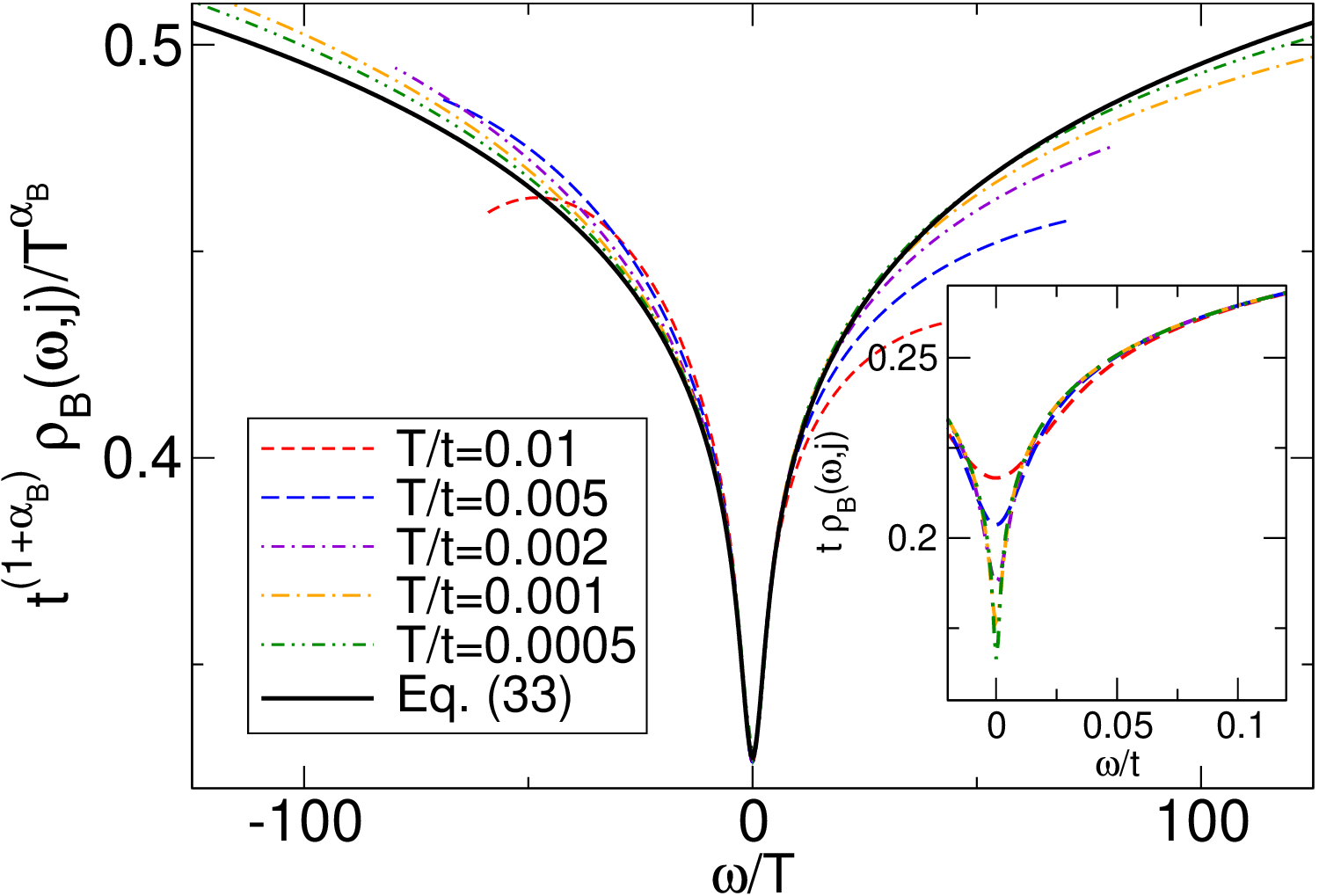}
   \caption{(Color online) Scaling plot of the local spectral function of the extended Hubbard 
     model for $n=3/4$, $U/t=0.5$, $V=V_{\rm o} = U/\sqrt{2}$ (for $n=3/4$), 
     $N=2^{14}$, $j=1$ and different $T$. The inset shows the unscaled spectral function
      as a function of $\omega$ for different $T$.}
   \label{fig3}
\end{center}
\end{figure}

\begin{figure}[t]
\begin{center}
   \includegraphics[width=0.7\linewidth,clip]{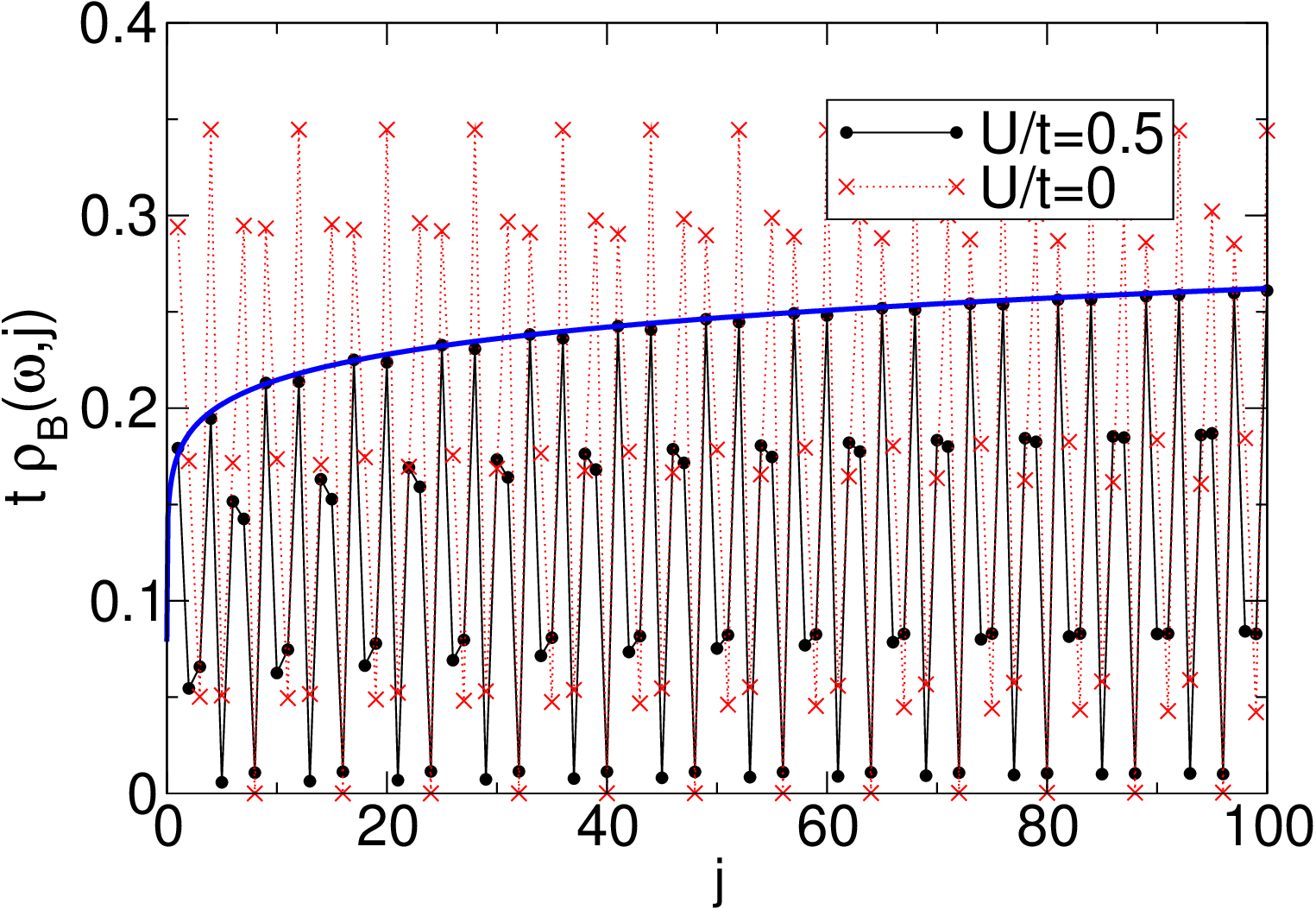}
   \caption{(Color online) Lattice site dependence of the spectral function of the extended Hubbard model 
     close to the boundary site $j=1$ for $n=3/4$, two different $U/t$, $V=V_{\rm o}=U/\sqrt{2}$ 
     (for $n=3/4$), 
     $N=2^{14}$, $T/t=10^{-3}$ and $\omega \approx 0$. The solid line is a power-law 
     fit to the envelope.}
   \label{fig4}
\end{center}
\end{figure}

To confirm scaling in $\omega/T$ at fixed $j$ as predicted in  Eq.~(\ref{eq:rhocompl}) 
we computed $\rho_{\rm B}$ for the parameters of Fig.~\ref{fig2} but with $j=1$ and for different 
$T$. By fitting $\rho_{\rm B}(\omega,j=1)$ as a function of $\omega$ for the smallest $T$ 
in the range $T \ll |\omega| \ll t$ we can extract a functional RG estimate of the boundary exponent 
$\alpha_{\rm B}^{\rm fRG}$. For the given parameters we obtain 
$\alpha_{\rm B}^{\rm fRG}=0.089$ in good agreement with the DMRG result
$\alpha_{\rm B}^{\rm DMRG}=0.095$ obtained from Eq.~(\ref{eq:alphaBdef}) and 
$K_c^{\rm DMRG}=0.840$ derived as explained above  \cite{Ejima05}. The scaling 
obtained with this $\alpha_{\rm B}^{\rm fRG}$ is shown in Fig.~\ref{fig3}. 
In the inset the unscaled data are displayed. The thick solid line is the prediction 
Eq.~(\ref{eq:rhocompl}) of the TL model with an open boundary, where we replaced $\alpha_{\rm B} 
\to \alpha_{\rm B}^{\rm fRG}$. The data nicely collapse on the TL model curve 
in the low-energy regime. 

We next take a closer look at the $j$ dependence of $\rho_{\rm B}$ at fixed $\omega$. 
As emphasized in the last 
section measuring the local spectral weight as a function of $j$ offers another possibility 
for a consistency check that the system under consideration is a LL: Eq.~(\ref{eq:rhocompl})
predicts power-law behavior of the envelope with exponent $2c=(1/K_{\rm c} - K_{\rm c})/4$.
Figure \ref{fig4} shows $\rho_{\rm B}$ for $n=3/4$, $U/t=0.5$,  
$N=2^{14}$, $T/t=10^{-3}$ and $\omega \approx 0$ as a function of $j$ (filled circles).
Here $\omega \approx 0$ refers to taking the eigenvalue of the finite system closest 
to $\omega=0$, which might be of order $1/N$ away from zero.    
We again tune $V$ to the optimal value $V_{\rm o}=U/\sqrt{2}$ (for $n=3/4$) 
providing the largest $\Delta$.
The spatial oscillations with frequency $2 k_{\rm F} = 3\pi/4$ are apparent. 
We fitted the envelope to a power law  $\sim j^{2 c^{\rm fRG}}$ and obtained 
$2 c^{\rm fRG} = 0.087$ in excellent agreement with
$2c^{\rm DMRG}=(1/K^{\rm DMRG}_{\rm c} - K^{\rm DMRG}_{\rm c})/4 = 0.088$. 
The power-law fit is shown as the thick solid line in Fig.~\ref{fig4}.
As mentioned above we control the different exponents only to leading order in the 
interaction. To this order the analytic expressions for $\alpha_{\rm B}$ and $2 c$ agree, 
as is apparent from Eqs.~(\ref{eq:expcdef}), (\ref{eq:alphaBdef}) and (\ref{eq:Kexp}).
The numerical values for $2 c^{\rm fRG}$ and $\alpha_{\rm B}^{\rm fRG}$ still differ 
by roughly 2\% as the RG produces higher than linear order terms in the different 
exponents as well.
We emphasize that for $V=0$, that is for the Hubbard model, in a similar plot 
no spatial suppression of the envelope of the spectral weight at small $j$ is visible. In fact, the 
envelope of the spectral weight at $\omega \approx 0$ even {\em increases} for $j$ approaching the 
boundary site $j=1$.    

A comparison of the data for $U/t=0.5$ (filled circles) and for $U/t=0$ (crosses) additionally 
presented in Fig.~\ref{fig4} shows that a {\em phase shift} $\xi$ of the spatial 
$2 k_{\rm F}$ oscillations appears which is {\em not} captured by the result 
for the TL model Eq.~(\ref{eq:rhocompl}). The latter was  derived using standard bosonization with
the boundary conditions on the continuum fields given after Eq.~(\ref{eq:gaussfields}). The 
deviation of the $U/t=0$ curve from Eq.~(\ref{eq:rhocompl0}) for $j \gtrapprox 20$ 
(splitting of degenerate values of the spectral weight) is a finite size 
effect; $\omega$ is not exactly zero but of order $1/N$ (eigenvalue closest to zero).     
The phase shift can be most easily identified from the observation that $\rho_{\rm B}$ vanishes 
on every eighth lattice site for $U/t=0$ (crosses), as it is supposed to according to 
Eq.~(\ref{eq:rhocompl0}) with $\omega=0$ and $2 k_{\rm F} =3 \pi/4$, but not so for 
$U/t=0.5$ (filled circles), where the same should hold according to Eq.~(\ref{eq:rhocompl}). 
The phase $\xi$ turns out to be {\em linearly} dependent on $U$ (for small $U/t$ and 
$V = V_{\rm o}(U)$). As our functional RG approximation scheme 
is controlled to this order, the appearance of $\xi$ is a reliable finding. Considering 
$T=0$ at different system sizes $N$ we 
furthermore verified that $\xi$ does not vanish for decreasing $1/N$. The phase shift is 
thus not a finite size effect. A phase shift as observed in the extended Hubbard model 
(with $V = V_{\rm o}(U)$) can be accounted for in bosonization by adding a 
{\em local single-particle forward scattering term} 
$W \delta(x) \, \partial_x \Phi_{\rm c}(x)$ to the Hamiltonian density Eq.~(\ref{eq:hamdens}). 
The phase shift is then given by $2 \pi K_{\rm c} W/v_{\rm c}$. To match the result of 
the extended Hubbard model (with $V = V_{\rm o}(U)$) $W$ has to be chosen $U$-dependent, 
in particular $W \sim U$ for small $U/t$. 
The STS and PES experiments always imply a spatial averaging. As the phase shift becomes 
illusive even after averaging over only a few lattice sites and as the main focus of the 
present paper is on relating theoretical spectral functions to experimental ones we here 
do not further investigate this issue. 

\begin{figure}[t]
\begin{center}
   \includegraphics[width=0.7\linewidth,clip]{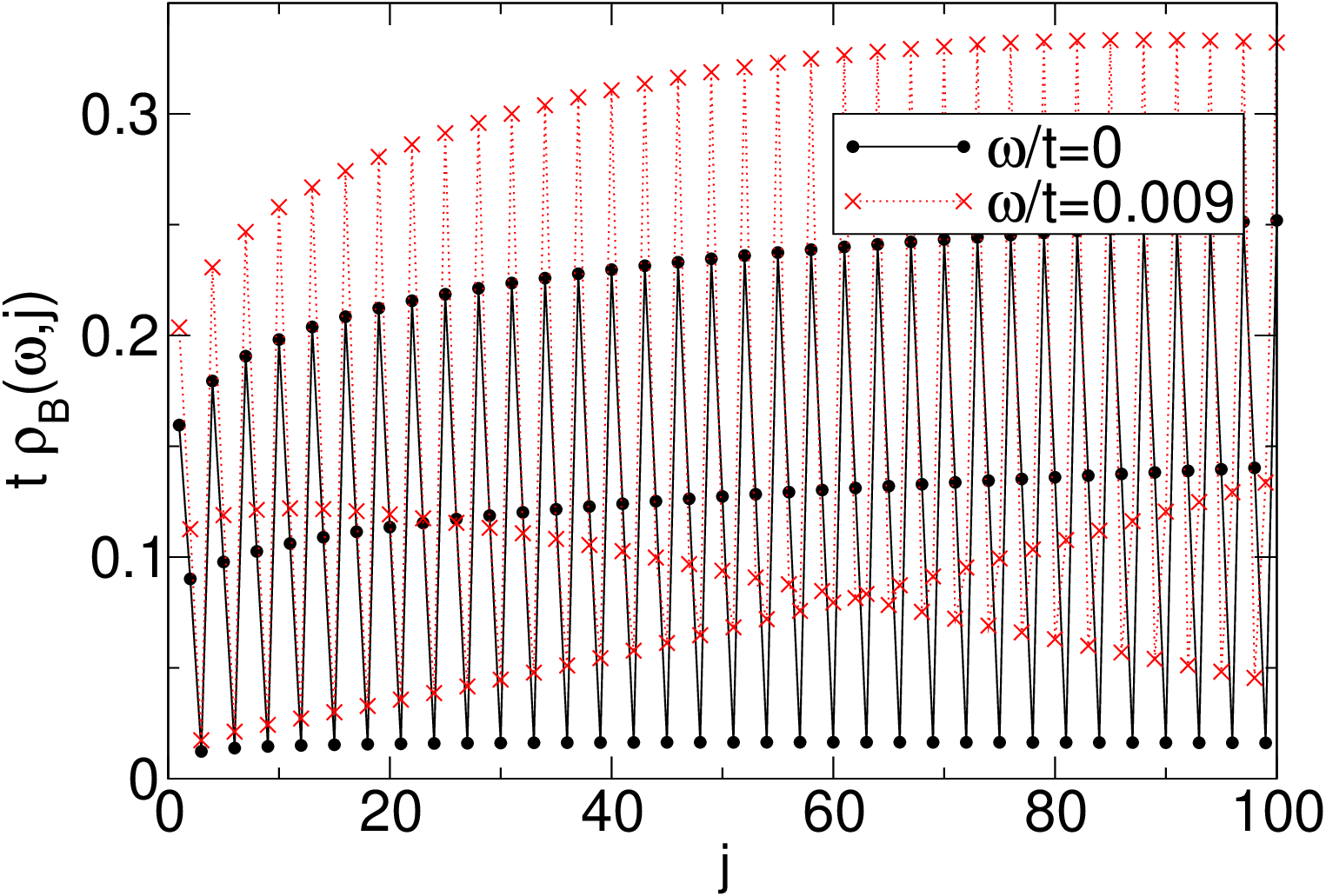}
   \caption{(Color online) Lattice site dependence of the spectral function of the extended Hubbard model 
     close to the boundary site $j=1$ for $n=2/3$, two different $\omega$, $U/t=0.5$, $V=V_{\rm o}=U$ 
     (for $n=2/3$), $N=2^{14}$, $T/t=10^{-3}$.}
   \label{fig5}
\end{center}
\end{figure}

Finally we compare the spatial dependence of $\rho_{\rm B}$ for two different $\omega$ in 
Fig.~\ref{fig5}. Apparently the frequency of the spatial oscillations depends on $\omega$ 
which is consistent with Eq.~(\ref{eq:rhocompl}). As this is more transparent for a 
'more commensurable' filling, the parameters of this figure are  $n=2/3$, $U/t=0.5$,  
$V=V_{\rm o}=U$ (for $n=2/3$), $N=2^{14}$ and $T/t=10^{-3}$.    

\section{Summary}
\label{sec:summary}

In the first part of this paper we have derived analytic expressions for the power-law  
behavior of the local spectral weight $\rho$ of the translational invariant TL model as well 
as the one with an open boundary $\rho_{\rm B}$ as a function of 
$\omega$, $T$ and $x$. The results  provide a variety of possibilities 
for consistency checks of experimental STS and PES data on 1d electron systems. The first is 
to show scaling of data for different $\omega$ and $T$ taken in the bulk or at fixed 
position close to the boundary onto the bosonization predictions Eqs.~(\ref{eq:rhotransinv}) 
and (\ref{eq:diveresedeffs}) (bulk) or Eq.~(\ref{eq:rhocompl}) (boundary). Scaling of bulk 
spectra was e.g.~demonstrated in Ref.~\cite{Blumenstein11}. If the same could be achieved 
for boundary spectra the {\em same} scaling function 
with $\alpha$ replaced by $\alpha_{\rm B}$ should appear if the system is a LL.
Experimentally showing this 
together with the required consistency of $\alpha$ and $\alpha_{\rm B}$ 
(for a spin-rotational invariant model with $K_{\rm s}=1$ both given by a
single number $K_{\rm c}$; see Eqs.~(\ref{eq:diveresedeffs}) and (\ref{eq:alphaBdef}))
would constitute a second highly nontrivial check that the studied system indeed is a LL. 
A consistency of bulk and boundary exponents within the experimental error bars  
(but not the entire scaling function) was achieved in Refs.~\cite{Bockrath99} 
and \cite{Blumenstein11} (also see below). A third consistency check is provided by the predicted 
spatial power-law behavior with the exponent $2c$ which again can be expressed 
solely in terms of $K_{\rm c}$ (see Eq.~(\ref{eq:expcdef})).    

It is often argued that the spatially oscillating contributions $\rho_{2 k_{\rm F}}$
and  $\rho_{-2 k_{\rm F}}$
to $\rho_{\rm B}$ with frequency $2 k_{\rm F}$ can be neglected when comparing to experimental 
spectra due to spatial averaging effects. Taking the numbers from Ref.~\cite{Blumenstein11} 
this is not apparent. In this experiment $k_{\rm F} \approx 5 \times 10^8/$m while the range 
of spatial averaging  is estimated as $\Delta x \approx 5 \times 10^{-9}\,$m, leading 
to $2 k_{\rm F} \Delta x=5$. This is not a very large number and one would thus conclude 
that $\rho_{2 k_{\rm F}}$ and  $\rho_{-2 k_{\rm F}}$ cannot be dropped. A quantitative 
picture of the averaging effects can easily be obtained 
by integrating Eq.~(\ref{eq:summe})  over an appropriate spatial range. We have 
explicitely verified that averaging over $\Delta x \approx 5 \times 10^{-9}\,$m 
does not significantly smear out the boundary ($x \omega/v_{\rm c}  \ll 1$)  and bulk 
($x \omega/v_{\rm c}  \gg 1$) exponents in $\omega$ at low temperatures (here $T=0$). 
In particular, this shows that the spatial resolution of the experiment is  
high enough to detect $\alpha_{\rm B}$ (as it is implicit to the analysis presented in 
Ref.~\cite{Blumenstein11}).   

In connection with the comparison of the experimental results on gold chains on a germanium 
surface of Ref.~\cite{Blumenstein11}
to the LL predictions one might be worried about two effects which are not included
in the TL model of Sect.~\ref{sec:scalfun}. One is Rashba spin-orbit interaction 
(SOI), which in a surface setup can become sizable. Along  the lines of Refs.~\cite{Governale04} 
and \cite{Schulz09} one can bosonize the 1d electron gas with Rashba SOI. An important 
effect is the appearance of {\em two} different Fermi velocities $v_{\rm F} (1 \pm \zeta)$ 
due to subband mixing and the SOI splitting \cite{Governale04,Schulz09}. Here $\zeta$ is a 
measure of the strength of 
the SOI. The local spectral function (with and without an open boundary) shows 
the same characteristics as a function of $\omega$ as discussed in Sect.~\ref{sec:scalfun}, 
but with modified exponents
\begin{eqnarray}
&&  \alpha^{\rm SOI} =  \alpha - \frac{\left(K_{\rm c}-1\right)^2}{2 \left(1 + K_{\rm c} \right)} \, \zeta 
+ {\mathcal O}(\zeta^2) , \\ 
&& \alpha_{\rm B}^{\rm SOI}  = \alpha_{\rm B} +  \frac{K_{\rm c}-1}{K_{\rm c} +1} \, \zeta 
+ {\mathcal O}(\zeta^2) .
\end{eqnarray}
Taking realistic numbers for the velocities it turns out that $\zeta \ll 1$ and the effects of Rashba 
SOI on the exponents are negligible. We note in passing that also the momentum resolved spectral function 
of a translational invariant LL is barely modified by SOI of realistic size \cite{Schulz10}.

The other issue is the observation of {\em four} electron branches 
(instead of two in the TL model) crossing the Fermi surface in PES measurements 
on the gold chains \cite{Meyer11}. The four branch situation can also be accounted for in 
bosonization \cite{Egger97,Kane97}. One then has to introduce even and odd pairs of charge and spin density 
bosons as well as the related $K$'s and velocities. Under the plausible 
{\em assumption} that only the even charge channel LL parameter is different from 
the noninteracting value 
$1$, that is $K_{\rm c,e} = K_{\rm c} <1 $, one finds the same power-law behavior for 
the local spectral functions as a function of $\omega$ as the ones given in 
Sect.~\ref{sec:scalfun} but with 
\begin{eqnarray}
&& \alpha= \frac{1}{8} \left( K_{\rm c} + \frac{1}{K_{\rm c}} -2 \right), \\ 
&& \alpha_{\rm B}=  \frac{1}{4} \left( \frac{1}{K_{\rm c}}-1 \right) .
\end{eqnarray} 
Interestingly taking these expressions would significantly improve the consistency of the 
experimentally determined  $\alpha$  and $\alpha_{\rm B}$ \cite{Blumenstein11}. From  
the measured value $\alpha=0.53$ of the bulk spectra one finds $\alpha_{\rm B}=1.27$ which 
nicely agrees to the measured value $\alpha_{\rm B}=1.20$. In particular, the agreement is 
improved compared to taking the two-branch expressions Eqs.~(\ref{eq:diveresedeffs}) 
and (\ref{eq:alphaBdef}) which gives  $\alpha_{\rm B}=1.43$ \cite{Blumenstein11}.  

In the second part of our paper we have shown that the behavior of $\rho_{\rm B}$ as a function 
of $\omega$, $T$ and $j$ close to an open boundary predicted by the 
bosonization solution of the TL model can indeed be found in an example of a  microscopic lattice 
model, namely the extended Hubbard model. Interestingly, a linear-in-$U$ phase shift $\xi$ not captured 
by standard open-boundary bosonization appears in the spatial oscillations of the spectral 
weight of the extended Hubbard 
model. The energy resolution required to access the low-energy LL regime strongly depends on the 
model parameters. Even for the fairly high energy resolution we can achieve within our approximate 
method, convincingly demonstrating LL power-law behavior and reliably extracting exponents requires 
fine-tuning of the parameters. Roughly speaking the crossover scale becomes small if the two-particle
interaction becomes too local. This suggests that to access the low-energy LL regime at a given 
experimental energy resolution one should consider systems with poor screening properties.

\noindent {\bf Acknowledgements:} We are grateful to J\"org Sch\"afer, Ralph Claessen, Kurt Sch\"onhammer 
and Christian Ast for very fruitful discussions. The numerical calculations were performed 
on the machines of the Max-Planck Institute for Solid State Research, Stuttgart. This work was 
supported by the DFG via the Emmy-Noether program (D.S.) and FOR 723 (S.A. and V.M.).

{}


\vspace*{.5cm}
\begin{thebibliography}{}

\bibitem{Voit95} 
  Voit J 1995 {\em Rep.~Prog.~Phys.} {\bf 58} 977 

\bibitem{Giamarchi03}
  Giamarchi T 2003  {\em Quantum Physics in One Dimension} (New York: Oxford 
  University Press)

\bibitem{Schoenhammer05} 
  Sch\"onhammer K 2005 in 
  {\em Interacting Electrons in Low Dimensions} ed. by D. Baeriswyl (Dordrecht: 
  Kluwer Academic Publishers) 

\bibitem{Grioni09} 
  For a review on the experimental status until 2009 see:
  Grioni M, Pons S and Frantzeskakis E 2009 {\em J.~Phys.: Condens.~Matter} {\bf 21} 
  023201  

\bibitem{Theumann67} 
  Theumann A 1967 {\em J.~Math.~Phys.} {\bf 8} 2460


\bibitem{Dzyaloshinskii73}
  Dzyaloshinskii IE and Larkin AI 1973 {\em Zh.~Eksp.~Teor.~Fiz.} {\bf 65} 411
  (english translation: 1974 {\em Sov.~Phys.-JETP} {\bf 38} 202)

\bibitem{Luther74}
  Luther A and Peschel I 1974 {\em Phys.~Rev.} B {\bf 9} 2911

\bibitem{Meden92} 
  Meden V and Sch\"onhammer K 1992 {\em Phys.~Rev.~} B {\bf 46} 15753

\bibitem{Schoenhammer93a}  
  Sch\"onhammer K  and Meden V 1993 {\em Phys.~Rev.~} B {\bf 47} 16205

\bibitem{Voit93} 
  Voit J 1993  {\em Phys.~Rev.~} B {\bf 47} 6740

\bibitem{Schoenhammer93b} 
  Sch\"onhammer K and Meden V 1993 {\it J. Electr.~Spectrosc.~Relat.~Phenom.} {\bf 62} 225

\bibitem{Haldane81} 
  Haldane FDM 1981 {\em J.~Phys.} C {\bf 14} 2585 

\bibitem{Solyom79} 
  S{\'o}lyom J 1979 {\em Adv.~Phys.} {\bf 28} 209

\bibitem{Kane92} 
  Kane CL and Fisher MPA 1992 {\em Phys.~Rev.} B {\bf 46} 15233 

\bibitem{Fabrizio95} 
  Fabrizio M and Gogolin AO 1995 {\em Phys.~Rev.} B {\bf 51} 17827

\bibitem{Eggert96}
  Eggert S, Johannesson H and Mattsson A 1996 {\em Phys.~Rev.~Lett.} {\bf 76} 1505

\bibitem{Grap09}
 Grap S and  Meden V 2009 {\em Phys.~Rev.} B {\bf 80} 193106

\bibitem{Ishii03} 
  Ishii H, Kataura H, Shiozawa H, Yoshioka H, Otsubo H, Takayama Y, Miyahara T, Suzuki S,
  Achiba Y, Nakatake M, Narimura T, Higashiguchi M, Shimada K, Namatame H and 
  Taniguchi M 2003 {\em Nature} {\bf 426} 540 

\bibitem{Hager05} 
  Hager J, Matzdorf R, He J, Jin R, Mandrus D, Cazalilla MA and Plummer EW 
  2005 {\em Phys.~Rev.~Lett.} {\bf 95} 186402

\bibitem{Wang06} 
  Wang F, Alvarez JV, Mo S-K, Allen JW, Gweon G-H, He J, Jin R, Mandrus D and H\"ochst H   
  2006  {\em Phys.~Rev.~Lett.} {\bf 96} 196403

\bibitem{Jompol09} 
  Jompol Y, Ford CJB, Griffiths JP, Farrer I, Jones GAC, Anderson D, Ritchie DA, Silk TW and 
  Schofield AJ 2009 {\em Science} {\bf 325} 597

\bibitem{Blumenstein11} 
   Blumenstein C, Sch\"afer J, Mietke S, Meyer S, Dollinger A, Lochner M, 
   Cui XY, Patthey L, Matzdorf R and Claessen R 2011 
    {\em Nature Physics} {\bf 7} 776

\bibitem{Meden99} 
  Meden V 1999 {\em Phys.~Rev.} B {\bf 60} 4571

\bibitem{Imambekov11} 
  Imambekov A, Schmidt TL and Glazman LI 2011 arXiv:1110.1374

\bibitem{Metzner11}
   Metzner W, Salmhofer M, Honerkamp C, Meden V and Sch\"onhammer K 2011 
   arXiv:1105.5289, to be published in Rev.\ Mod.\ Phys.

\bibitem{Schuricht11} 
  Schuricht D, Essler FHL, Jaefari A and Fradkin E 2008 {\em Phys.~Rev.~Lett.} {\bf 101} 086403; 
  2011 {\em Phys.~Rev.} B {\bf 83} 035111; Schuricht D 2011  {\em Phys.~Rev.} B {\bf 84} 045122

\bibitem{Bockrath99} 
  Bockrath M, Cobden DH, Lu J, Rinzler AG, Smalley RE, Balents L and 
  McEuen PL 1999 {\em Nature} {\bf 397} 598

\bibitem{Rodin10} 
  Rodin AS and Fogler MM 2010 {\em Phys.~Rev.~Lett.} {\bf 105} 106801

\bibitem{Eggert00} 
  Eggert S 2000 {\em Phys.~Rev.~Lett.} {\bf 84} 4413

\bibitem{Preus94} 
  Preus R, Muramatsu A, von der Linden W, Dieterich P, Assaad FF and Hanke W 
  1994 {\em Phys.~Rev.~Lett.} {\bf 73} 732

\bibitem{Benthien05} 
  Benthien H 2005 PhD thesis, Universit\"at Marburg

\bibitem{Benthien04} 
  Benthien H, Gebhard F and Jeckelmann E 2004 {\em Phys.~Rev.~Lett.} {\bf 69} 256401

\bibitem{Abendschein06}  
  Abendschein A and Assaad FF 2006 {\em Phys.~Rev.} B {\bf 73} 165119 

\bibitem{Jeckelmann11} Jeckelmann E 2011 arXiv:1111.6545 and this special issue

\bibitem{Enss05} 
  Enss T, Meden V,  Andergassen S, Barnab\'e-Th\'eriault X, Metzner W and Sch\"onhammer K 2005
  {\em Phys.~Rev.} B {\bf 71} 155401  

\bibitem{Andergassen06}
  Andergassen S, Enss T, Meden V, Metzner W, Schollw\"ock U and Sch\"onhammer K 2006
  {\em Phys.~Rev.} B {\bf 73} 045125  

\bibitem{Essler05} 
  Essler FHL, Frahm H, G\"ohmann F, Kl\"umper A and Korepin VE 2005 
  {\em The One-Dimensional Hubbard Model} (Cambridge: Cambridge University Press)

\bibitem{Schulz90} 
  Schulz HJ 1990 {\em Phys.~Rev.~Lett.} {\bf 64} 2831

\bibitem{Ejima05}
  Ejima S, Gebhard F and Nishimoto S 2005 
  {\em Europhys.~Lett.} {\bf 70} 492

\bibitem{Meden00} 
  Meden V, Metzner W, Schollw\"ock U, Schneider O, Stauber T, and Sch\"onhammer K 2000
  {\em Eur.~Phys.~J.} B  {\bf 16} 631 

\bibitem{Andergassen04}
  Andergassen S, Enss T, Meden V, Metzner W, Schollw\"ock U, and Sch\"onhammer K
  2004 {\em Phys.~Rev.} B {\bf 70} 075102 

\bibitem{Governale04} 
  Governale M and Z\"ulicke U 2004 {\em Solid State Com.} {\bf 131} 581

\bibitem{Schulz09} 
  Schulz A, De Martino A, Ingenhoven P and Egger R 2009
  {\em Phys.~Rev.} B {\bf 79} 205432

\bibitem{Schulz10}
   Schulz A, De Martino A and Egger R 2010
  {\em Phys.~Rev.} B {\bf 82} 033407

\bibitem{Meyer11}
  Meyer S, Sch\"afer J, Blumenstein C, H\"opfner P, Bostwick A, 
  McChesney JL, Rotenberg E, Claessen R 2011 {\em Phys.~Rev.} B {\bf 83} 121411(R)

\bibitem{Egger97}
  Egger R and Gogolin AO 1997 {\em Phys.~Rev.~Lett.} {\bf 79} 5082

\bibitem{Kane97}
  Kane C, Balents L and Fisher MPA 1997 {\em Phys.~Rev.~Lett.} {\bf 79} 5086

\end{thebibliography}
\end{document}